\documentclass[%
reprint,
superscriptaddress,
amsmath,amssymb,
aps,
pre,
]{revtex4-1}

\usepackage{dcolumn}
\usepackage{float}
\usepackage{bm}
\usepackage{comment} 
\usepackage{hyperref}
\usepackage{amssymb,amsmath,graphicx,array,color,amsthm}
\usepackage{caption}
\usepackage[skip=-5pt]{subcaption}
\usepackage{amscd}
\usepackage{multirow}
\usepackage{float}
\def\bfk{{\bm{k}}}
\def\rmd{\,\mathrm{d}\,}
\def\k{{\bf{k}}}
\newcommand{\Sec}[1]{{\textit{#1.---}}}

\begin{document}
\preprint{AIP/123-QED}

\title{Strong and weak wave turbulence regimes in Bose-Einstein condensates}

\author{Ying Zhu}
\email{zhuying0325@gmail.com}
\affiliation{Universit\'{e} C\^{o}te d'Azur, CNRS, Institut de Physique de Nice {\color{black}(INPHYNI)}, 17 rue Julien Lauprêtre, 06200 Nice, France}
\author{Giorgio Krstulovic}
\affiliation{Universit\'{e} C\^{o}te d'Azur, CNRS, Institut de Physique de Nice {\color{black}(INPHYNI)}, 17 rue Julien Lauprêtre, 06200 Nice, France}
\affiliation{Universit\'{e} C\^{o}te d'Azur, Observatoire de la C\^{o}te d'Azur, CNRS, Laboratoire Lagrange, Boulevard de l'Observatoire CS 34229 -- F 06304 Nice Cedex 4, France}
\email{krstulovic@oca.eu}
\author{Sergey Nazarenko}
\affiliation{Universit\'{e} C\^{o}te d'Azur, CNRS, Institut de Physique de Nice {\color{black}(INPHYNI)}, 17 rue Julien Lauprêtre, 06200  Nice, France}

\begin{abstract}
When a turbulent Bose-Einstein condensate is driven out-of-equilibrium at a scale much smaller than the system size, nonlinear wave interactions transfer particles towards large scales in an inverse cascade process.
In this work, we numerically study wave turbulence in a three-dimensional Bose-Einstein condensate in forced and dissipative inverse cascade settings. We observe that when the forcing rate increases, thereby increasing the particle flux, the turbulence spectrum gradually transitions from the weak-wave Kolmogorov-Zakharov cascade to a critical balance state characterized by a range of scales with balanced linear and nonlinear dynamic timescales.
Further forcing increases lead to a coherent condensate component superimposed with Bogoliubov-type acoustic turbulence. The role of vortices in such a strongly forced state is marginal, which makes this new state distinct from the strongly turbulent state composed of a tangle of quantized vortex lines. We then use our predictions and numerical data to formulate a new out-of-equilibrium equation of state for the 3D inverse cascade. 
\end{abstract}
\maketitle

\Sec{Introduction} Turbulence is a central example of strongly non-equilibrium systems whose state is determined by an energy (or another invariant) flux through scales. When forcing and dissipation scales are well separated, an inertial range of scales emerges, where the dynamics is driven by the nonlinear interactions only. The central concept of turbulence is the idea of cascade, originally proposed by Richardson in the case of hydrodynamic turbulence. In 1941, Kolmogorov pushed this concept further by postulating that in the limit of an infinite inertial range, the energy dissipation rate $\varepsilon$ is finite, necessarily equal to the energy flux, and it is the only dimensional parameter involved in the turbulent energy cascade. A simple dimensional analysis then suffices to fix scaling exponents, which leads to the famous Kolmogorov energy spectrum $E_k\propto \varepsilon^{2/3}k^{-5/3}$, with $k$ the wavevector \cite{frischTurbulenceLegacyKolmogorov1995}. Many other turbulent systems exist where more than one dimensional parameter comes into play, so scaling exponents cannot be fixed so simply. Most typical examples are magnetohydrodynamics and nonlinear wave systems such as internal and inertial waves in the oceans and atmospheres, capillary-gravity waves on the surface of liquids, or Bose-Einstein condensates \cite{GALTIER_NAZARENKO_NEWELL_POUQUET_2000,Galtier_WeakInertialwaveTurbulence_2003,lvovHamiltonianFormalismGarrettMunk2001,Falcon_ExperimentsSurfaceGravity_2022}. Fortunately, when nonlinearity is small, the Weak Wave Turbulence Theory (WWTT) predicts the whole spectrum, including universal dimensionless pre-factors \cite{nazarenko2011wave}. When nonlinearity becomes strong, only phenomenological arguments can be provided in rare cases.
How a system transitions from the weak to the strong wave turbulence (WT) regimes remains an open question.

In this Letter we study the weak-to-strong transition in turbulence of Bose-Einstein condensates (BECs).  BECs provide an excellent experimental platform for studying turbulence in general and WT in particular because of the versatility of optical techniques in creating experimental setups and diagnostics. 
Several types of BEC turbulence setups in two and three dimensions have been realized by various types of forcing, from shaking the BEC  trap to dragging repulsive potentials through the system, thereby generating quantized vortices and/or dispersive waves \cite{Bagnato2009,Bagnato2011,Kwon,navon2016emergence,navon2019synthetic,Bagnato2020,galkaEmergenceIsotropyDynamic2022,karailievObservationInverseTurbulentWave2024,Johnstone,Gauthier}.
Recently, experimentalists using a shaking trap have succeeded in creating turbulent BECs incorporating a synthetic small-scale dissipation mechanism \cite{navon2019synthetic,galkaEmergenceIsotropyDynamic2022,karailievObservationInverseTurbulentWave2024}.  Introducing such a synthetic dissipation is an important new element as it allows for producing stationary  BEC turbulence and, therefore, testing theoretical predictions for the statistically steady state. In particular, a stationary WT state corresponding to a direct local
cascade of energy from low momenta (where it was forced) to high momenta (where it was dissipated) was reported in 
\cite{navon2019synthetic}. The resulting spectrum was explained theoretically
and confirmed numerically in \cite{zhu2023direct}. The latter paper has also derived and simulated the stationary inverse WT cascade directed from a source at high momenta to a dissipation range at low momenta -- a setup yet to be implemented experimentally. The present work extends this study for a very wide range of forcing amplitudes, including cases far beyond the WWTT applicability. 

To model BECs,  we use the Gross-Pitaevskii equation (GPE) for the complex wave function $\psi({\bf x},t)$,
\begin{equation}
 i\frac{\partial\psi   }{\partial t}= -\alpha \nabla^2\psi +\beta |\psi|^2\psi,
 \label{GPE}
\end{equation} 
where $\alpha={\hbar}/{2m}$, $\beta=g/\hbar$
and $g=4\pi \hbar^2 a_s /m$, with $\hbar$ the reduced Planck constant, $m$ the boson mass, and $a_s$ the $s$-wave scattering length \cite{pitaevskii2003bose}.
 For simplicity, we consider a triply periodic cube of side $L$
and volume $V = L^3$. The GPE \eqref{GPE} conserves the total number of particles and energy per unit volume,
\begin{eqnarray}
N &=&\frac{1}{L^3}\int |\psi(\mathbf{x}, t)|^2 \rmd\mathbf{x} ,\label{eq:cons-N} \\
H&=&\frac{
\hbar}{L^3}\int \left[ \alpha|\nabla \psi(\mathbf{x}, t)|^2 + \frac{\beta}{2}|\psi(\mathbf{x}, t)|^4 \right] \rmd\mathbf{x} \, \label{eq:cons-H},
\end{eqnarray}
respectively. In a turbulent setting, energy is transferred towards small scales (direct cascade), whereas particles exhibit an inverse cascade towards large scales. Most of the turbulent BEC experiments have studied the direct energy cascade \cite{navon2019synthetic,galkaEmergenceIsotropyDynamic2022}, 
whereas inverse particle turbulent transfers have been recently observed both in 2D and 3D experiments \cite{
karailievObservationInverseTurbulentWave2024, moreno2025observation}. (The inverse particle cascade should not be confused with the inverse energy cascades and clustering in the 2D BEC quantized vortex turbulence studied in \cite{Johnstone,Gauthier}.)

In this Letter, we focus on the inverse particle cascade in 3D. In particular, we study how the system transitions from weak to strong wave turbulence as the particle flux per unit volume $|Q_0|$ is increased. The advantage of this setting is that the WWTT provides an exact and simple analytical prediction, unlike the 3D direct energy cascade or the cascades in 2D BECs that present certain mathematical challenges.

\Sec{Weak Wave Turbulence and Critical Balance Predictions}
As usual in turbulence, let us start with dimensional considerations. The main observable in the turbulent cascade is the waveaction spectrum, $n_{\bf k}(t)\equiv n({\bf k},t) = \frac{V}{(2\pi)^3}\langle |\hat \psi_{\bf k}(t)|^2 \rangle$, 
where $\hat{\psi}_{\bf k}=\frac{1}{L^3}\int \psi(\mathbf{x})e^{i\k \cdot\mathbf{x}}\rmd \mathbf{x}$ is the Fourier transform of $\psi(\mathbf{x})$, and the brackets denote the average over fluctuations. Note  that $n_{\bf k}$ is dimensionless, and from Eq.~\eqref{GPE}, $[\alpha]=\rm{length}^2/\rm{time}$ and $[\beta]=\rm{length}^3/\rm{time}$. As $[Q_0]=\rm{length}^{-3}\rm{/time}$, then by matching dimensions we have
\begin{equation}
  n_{\bf k} = C \left(\frac{|Q_0|\beta^5}{\alpha^6}\right)^y \left(\frac{\alpha}{\beta k}\right)^x,
  \label{eq:gen_spec}
\end{equation}
where $C$  is a dimensionless constant, the two terms in parentheses are independent dimensionless combinations, and we assumed a power-law dependence on these parameters with $x$ and $y$ being {\em a priori} arbitrary exponents. 
Here, $x$ fixes the dependence of the waveaction spectrum on scales, whereas $y$ determines the behavior of the so-called out-of-equilibrium equation of state (EoS) \cite{dogra2023universal,zhu2024turbulence}. Indeed, the spectrum's amplitude dependence on the flux $Q_0$ is analogous to the 
dependence of gas density on the temperature at thermal equilibrium (e.g., at constant pressure). The WWTT predicts the Kolmogorov-Zakharov (KZ) spectrum that fixes all the exponents and the universal dimensionless constant $C$~\cite{zhu2023direct}:
 \begin{equation} \label{eq:ic}
  n_{ k} =C_{\rm i} \frac{\alpha^{1/3}}{\beta^{2/3}}|Q_0|^{1/3} k^{-7/3}\,,\;\; C_{\rm i}\approx 7.5774045\times 10^{-2}. 
 \end{equation}
 This spectrum is an exact solution of the four-wave kinetic equation associated with the GPE, 
 which is the central equation of the WWTT obtained under the assumptions of small amplitudes and random phases of waves. Accordingly, the constant $C_{\rm i}$ is given by an analytical expression, the approximate value of which is provided above. It is a WT analog of the Kolmogorov constant in the famous Kolmogorov spectrum of classical hydrodynamic turbulence~\cite{frischTurbulenceLegacyKolmogorov1995}.
 The KZ spectrum \eqref{eq:ic} was found to be in excellent agreement with numerical simulations \cite{zhu2023direct}. 
 
 When the nonlinearity is strong, phases might be strongly correlated, and the spectrum could be guessed by invoking the critical balance (CB) assumption where linear and nonlinear timescales 
 are balanced at each scale within a certain range of scales, a concept introduced in~\cite{GS1995} in the context of the magnetohydrodynamic turbulence.   In fact, approaches similar in spirit were independently proposed in various physical contexts (see examples in \cite{NAZARENKO_SCHEKOCHIHIN_2011}) with the first such theory developed in the context of water waves, which led to the famous Phillips spectrum \cite{Phillips_1958}. 
 The appearance of CB can often be understood as an effect of a set of coherent structures with a range of sizes generated in the system (possibly tightly packed in physical space). For example, the Phillips spectrum is often attributed to sharp-crested Stokes waves, which are limited to a critical amplitude by wave breaking \cite{Phillips_1985}. It is natural to think that in the GPE system, such CB states could be characterized by a set of densely packed solitonic structures of various sizes. 
 For GPE, the CB argument was first introduced in \cite{proment2009energy}, and here we will revise and refine it. We start with the GPE \eqref{GPE} written in Fourier space
 \begin{equation}
  i\frac{\partial\hat{\psi}_{\bf k}  }{\partial t}=\omega_k\hat{\psi}_{\bf k}  +\beta \sum_{ 123}\hat{\psi}_{ 1}^* \hat{\psi}_{2}\hat{\psi}_{ 3}\delta^{\bfk 1}_{23},\, {\rm with}\, \omega_k=\alpha k^2, \, k=|\k|,
  \label{Eq:NLSFourier}
\end{equation} 
where $\sum_{ 123}$ stands for a sum over all wave vectors $\k_1,\k_2,\k_3$, $\psi_{\bf i} = \psi(\k_{i})$, and 
 $\delta^{\bfk 1}_{23}$
 is a Kronecker delta: $1$ if $\k+\k_1=\k_2+\k_3$, $0$ otherwise. Note that for both WWTT and CB, the leading-order nonlinear effect in \eqref{Eq:NLSFourier} is provided by the diagonal terms in the sum, i.e., those with $\k =\k_2, \k_1 =\k_3 $ and $\k =\k_3, \k_1 =\k_2 $. They lead to a nonlinear frequency shift $\omega_k \to \omega_k + 2\beta N$, but do not affect the evolution of $|\hat{\psi}_{\bfk}| $ and, therefore, do not lead to any spectral redistribution.  They can be eliminated by passing into the rotating frame, $\hat{\psi}_{\bf k}  \to e^{-2i\beta Nt}\hat{\psi}_{\bf k}$. Applying the CB argument to \eqref{Eq:NLSFourier}, we find $\omega_k |\hat{\psi}_{\bf k}| \sim \beta |\hat{\psi}_{\bf k}|^3 (k/k_{\rm cor})^6$ where $k_{\rm cor}$ is the correlation distance in $k$-space (it is vanishingly small in the WWTT). The $(k/k_{\rm cor})^6$-term estimates the number of modes interacting with mode $k$. 
  With $n_{\bf k} = (L/2\pi)^3 \langle |\hat{\psi}_{\bf k}|^2 \rangle $, this leads to
$  n_{\bf k} \sim 
\frac{\alpha}{\beta } \frac{k_{\rm cor}^6 }{k^4} \frac{L^3}{(2\pi)^3}. 
$
In the limit $L\to \infty$, this expression should be independent of $L$. If we additionally postulate that $k_{\rm cor}$ is independent of $k$ then the most general expression for it (leading to an $L$-independent spectrum) is $k_{\rm cor} \sim \left(\frac{|Q_0| \beta^5 }{\alpha^6}\right)^{y/6} 
\left(\frac{2\pi \alpha }{L \beta}\right)^{1/2}$. This finally gives
\begin{equation}
  n_{\bf k} 
  = C \left(\frac{|Q_0|\beta^5}{\alpha^6}\right)^y 
  \left(\frac{\alpha}{\beta k}\right)^4
  \label{eq:CB}
\end{equation}
i.e., Eq.~\eqref{eq:gen_spec}
with $x=4$. Note that this value is a consequence of our assumption that $k_{\rm cor}$ is $k$-independent.
Unfortunately, the CB argument is too basic to be able to predict the exponent $y$, i.e., to determine how 
$k_{\rm cor}$ scales with the flux. We will test the scaling $x=4$ and extract the value of $y$ using the numerical simulations.

\Sec{Numerical Simulations}
We perform numerical simulations of the forced-dissipated GPE using the standard massively-parallel pseudo-spectral code FROST \cite{KrstulovicHDR} with a fourth-order Exponential Runge-Kutta temporal scheme (see \cite{zhu2023direct}).
We adimensionalize  the GPE by measuring length in units of $\beta/\alpha$, time in units of $\beta^2/\alpha^3$,
 and $\psi$ in units of $(\alpha/\beta)^{3/2}$,
which amounts to setting $\alpha=\beta=1$ in Eq. \eqref{GPE}.
We set $L=2\pi$, which means that the nondimensional grid spacing in Fourier space is $\Delta k =1$, and use grids of $N_p^3$ collocation points, with $N_p=512$. 
We add a forcing term $F_{\bf k} (t) $ and a dissipation term $-D_{\bf k}\hat{\psi}_{\bf{k}} (t)$ to the RHS of the Fourier-space GPE \eqref{Eq:NLSFourier}.
To obtain an inverse cascade, the forcing term is supported on a narrow band at high $k$'s around the forcing wavenumber $k_{\rm f}$ for $k \in [k_{\rm f}-1, k_{\rm f}+1]$,
and it is given by
$\mathrm{d}F_{\bf k} (t)=f_0\mathrm{d}{W}_{\bf k}$, 
where {$W_{\bf k}$} is the Wiener process, and the parameter $f_0$ controls the amplitude of the forcing. 
Dissipation is implemented via hypoviscosity at the largest scales ($0 < k \le k_{\rm L} \ll k_{\rm f}$), taking the form $D_{\bf k} = D_{\rm L} k^{-2r}$, and via hyperviscosity at small scales (high wavenumbers) beyond the forcing range ($k_{\rm f}+1 < k \le k_{\rm max}$), where $D_{\bf k} = D_{\rm R} k^{2s}$.
Additionally, the condensate mode ($k=0$) is dissipated at a constant rate $D_0$. 
We optimize the parameters of forcing and dissipation to enlarge the inertial range for a fixed resolution, ensuring that the simulations are well-resolved and minimizing bottlenecks at the dissipation scales. All the numerical parameters are listed in Appendix \ref{append:a}. 
 We perform simulations with different values of the forcing strength $f_0$ in order to study the transition from the weak to the strong WT regimes.
The $k$-space  particle flux $Q(k)$ is computed directly from the RHS of the GPE \eqref{GPE} as in \cite{zhu2023direct} (details of this calculation method are provided in Appendix \ref{append:b}). We evaluate $Q_0$ (the rate of particle loss per unit volume and unit time through low-k dissipation) by measuring the value of $Q(k)$ in the inertial range where it is approximately constant.
Note that $f_0$ and $|Q_0|$ are monotonically related, so we discuss our results in terms of the strength of the particle flux. Finally, our numerical measurements are performed when a statistically steady state is reached.

\begin{figure}[h!]
  \centering
  \includegraphics[scale=0.95]{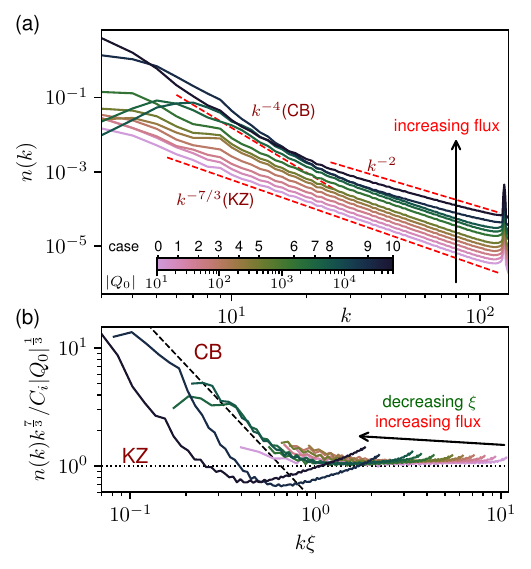}
  \caption{\label{fig:direct} Stationary spectra and scaling regimes across different flux strengths. (a) Variation of the waveaction spectra $n(k)$ for all simulation cases (cases 0--10). The arrow indicates the direction of increasing particle flux $|Q_0|$, with red dashed lines showing the predicted power-law scalings for the Kolmogorov-Zakharov (KZ, $k^{-7/3}$), critical balance (CB, $k^{-4}$), and thermal ($k^{-2}$) regimes.
(b) Compensated spectra $n(k)k^{7/3}/(C_{\rm i} |Q_0|^{1/3})$ as a function of the normalized wavenumber $k\xi$. The collapse of the data toward the horizontal dotted line at unity demonstrates the validity of the KZ prediction in the inertial range across varying flux strengths. Note: The coefficients $\alpha$ and $\beta$ are omitted in this and subsequent figures, as $\alpha = \beta = 1$ in all simulations.}
  \end{figure}

Figure \ref{fig:direct}(a) shows the stationary waveaction spectrum $n(k)$ for values of the flux $Q_0$ varying over almost five orders of magnitude. The respective spectra compensated by the KZ prediction \eqref{eq:ic} and the wavenumber compensated by the healing length
$\xi=\sqrt \alpha /\sqrt{\beta  N}$ 
are shown in Fig.~\ref{fig:direct}(b). As noted in the numerical methods, we set $\alpha = \beta = 1$ throughout; for clarity, these parameters are omitted from the figure axes and labels hereafter.
We observe that for lower values of $|Q_0|$ there is almost perfect agreement with the WWTT KZ spectrum prediction \eqref{eq:ic} -- with both the slope and the prefactor constant of this solution confirmed (as previously  seen in \cite{zhu2023direct}). Naturally, the range at which this agreement holds shrinks when the flux $|Q_0|$ increases, and the weak nonlinearity assumption fails at  $k\xi \lesssim 1$. This is natural because the healing length $\xi$ is defined as the scale where the linear and the nonlinear terms in the GPE are of a similar size. As scales $k\xi \lesssim 1$, the $k ^{-4}$ CB scaling \eqref{eq:CB} appears when the flux is moderately strong such that $k_\xi =1/\xi $ still lies within the inertial range. Note, however, that CB can only be observed as an intermediate state for moderately high fluxes.
Thus, CB is not the \emph{ultimate} strong wave turbulence regime. 
Indeed, for the highest particle flux, we observe a very steep spectrum at low $k$'s and close to a $k^{-2}$-spectrum at large wavenumbers. We will later relate this behavior to the emergence of a quasi-condensate with a thermal component induced by the forcing.
Condensation occurs when the infrared dissipation becomes insufficient for absorbing the particle flux  before it reaches the ground state.
The flux functions $Q(k)$ for all runs are provided in Fig.~\ref{fig:flux} of Appendix \ref{append:b}.
The figure illustrates that the flux at the lowest non-zero modes, $k=1$, is two-to-three orders of magnitude higher in runs with the condensation (cases 9 and 10) than in the rest of the runs where there was no condensation.
When the  spectrum pileup occurring at low wavenumbers starts exceeding a certain value, the interaction with these wavenumbers becomes dominant and this changes the dynamics to the acoustic type. Namely, the condensate affects, nonlocally in
the $k$-space, the dynamics of waves with higher wavenumbers: the dominant process is now such that one of the
four wavenumbers in the interacting quartet belongs to the low-$k$ condensate. This fact invalidates the locality
of the four-wave interaction property used by both the four-wave WWTT and the CB theories.

In order to quantitatively study the transition from the WWTT to the strong  wave turbulence, we make use of the spatio-temporal Fourier transform (STFT) of the wavefunction, which is obtained by performing the Fourier transform in both physical-space coordinate and in time over a finite (but sufficiently large) time window. STFT has become a standard tool in wave turbulence systems for checking if the system is in the weakly nonlinear regime: in this case STFT is narrowly concentrated in the proximity of the linear-wave dispersion relation $\omega_k = \alpha k^2$, i.e. the width of the frequency broadening of STFT spectrum at a fixed $k$, $\delta \omega(k)$, must remain much less than
$\omega_k$. 
Moreover, the CB and strong wave turbulence regimes are characterized by $\omega_k\sim \delta\omega_k$ and $\omega_k < \delta\omega_k$, respectively.

\begin{figure*}
    \centering
    \includegraphics[scale=0.95]{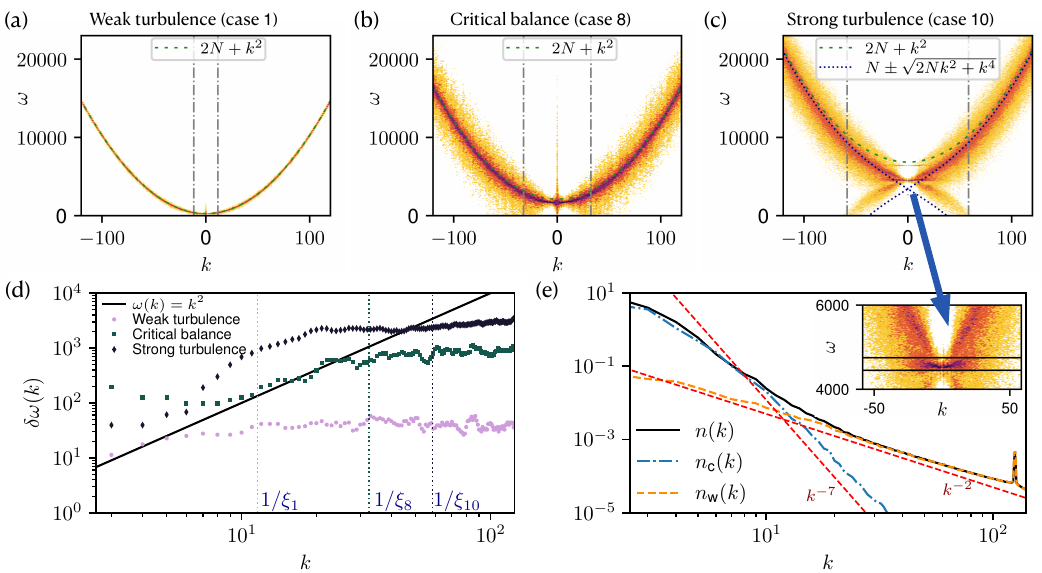}
    \caption{\label{fig:st} 
    Spatio-temporal analysis and spectral decomposition.
(a)–(c) Representative spatio-temporal Fourier transform spectra illustrating the transition across three flux regimes: weak turbulence (case 1), critical balance (case 8), and strong turbulence (case 10). The vertical dash-dotted lines indicate the characteristic scale $k = 1/\xi$ for each case.
(d) Nonlinear frequency broadening $\delta\omega(k)$ as a function of $k$ for the three representative cases shown in (a)–(c); the solid black line represents the relation $\delta\omega = \omega_k = k^2$, marking the critical balance regime where linear and nonlinear time scales are comparable.
(e) Spectral decomposition for the strong-flux case into the condensate component $n_{\rm c}(k)$ and the thermal component $n_{\rm w}(k)$. The dashed lines indicate the respective power-law fits: the $k^{-7}$ scaling for the condensate and the $k^{-2}$ Bogoliubov scaling for the thermal distribution.}
    \end{figure*}

Representative examples of the STFT spectral plots are shown in Figs.~\ref{fig:st}(a)--(c) and the measurements of the nonlinear frequency broadening $\delta \omega(k)$ (defined here as the width corresponding to $80\%$ of the area under the curve) are in Fig.~\ref{fig:direct}(d).
  We can see that in both the weak (a) and the intermediate flux cases (b), the spectrum is distributed closely to the dispersion relation 
  $\omega =\alpha k^2 +2\beta N$, with  $2\beta N$ being the nonlinear shift. The nonlinear broadening displayed in Fig.~\ref{fig:st}(d) shows that for the lowest flux case the weak nonlinearity condition for applicability of weak WT, $\delta \omega(k) \ll \omega_k$  is realized over a broad inertial range of wavenumbers between the large-scale dissipation and the forcing scales, $3 \lesssim k \lesssim 100$, the range where the inverse cascade KZ prediction is clearly observed. For the intermediate flux case, the range where $\delta \omega(k) \ll \omega_k$ is also present, but now in a narrower range, $30 \lesssim k \lesssim 100$. KZ spectrum is still seen in this range.
  On the left of it, in the range $10 \lesssim k \lesssim 30$, we observe $\delta \omega(k) \sim \omega_k$, which is, by definition, the CB state. Accordingly, the CB spectrum is indeed realised in this range. The transition between the KZ and the CB scaling occurs approximately at the healing length wavenumber, $k_\xi \sim 30$. The strong WT case deserves more explanation.

\begin{figure}
    \centering
    \includegraphics[scale=1]{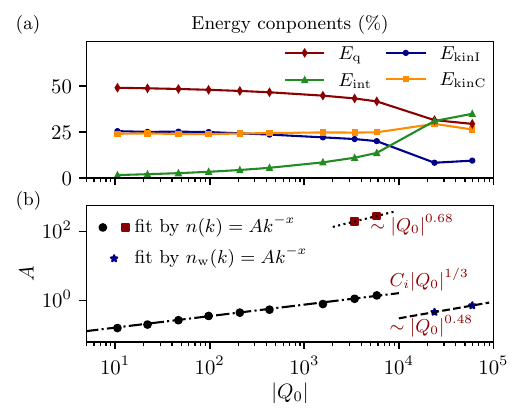}
    \caption{ Energy partitioning and equation of state for varying particle flux $|Q_0|$. (a) Percentages of energy components for the incompressible kinetic energy $E_{\text{kinI}}$, compressible kinetic energy $E_{\text{kinC}}$, quantum energy $E_q$, and internal energy $E_{\text{int}}$, respectively. (b) Fitted spectral amplitude $A$ and its scalings covering the weak, medium, and strong-flux regimes. The circles represent the amplitude $A$ obtained from the power-law fit $n(k) = Ak^{-x}$ in the weak turbulence regime at $k > 1/\xi$ (cases 0--8). For the intermediate cases 7 and 8, squares denote a local fit in a narrow large-scale range ($k < 1/\xi$) within the inertial zone. The stars represent the fit for the thermal component $n_{\rm{w}}(k)$ specifically for the high-flux cases 9 and 10.}
    \label{fig:eos}
\end{figure}

  \Sec{Emergence of Condensate and Acoustic Turbulence}
  When the flux is very intense, the CB assumption is no longer fulfilled and the system transitions to a new regime of  strong wave turbulence, exhibiting a very steep spectrum at low $k$'s (Fig.~\ref{fig:direct}). For this case, the STFT in Fig.~\ref{fig:eos}(c) displays a behavior completely different from a broadened 
  $\omega_k=\alpha k^2+2\beta N$ curve. This state is characterized by the emergence of a strong coherent condensate component around $k=0$. The condensate affects, nonlocally in the $k$-space, the dynamics of waves with higher wavenumbers, thereby invalidating the locality property used by both the WWTT and the CB theories. The effect of the condensate takes the form of making the wave perturbations behave as sound characterized by the well-known Bogoliubov dispersion relation,
  \begin{equation}
  \omega_B(k) =  \beta|\psi_0|^2 \pm \sqrt{2\alpha\beta |\psi_0|^2 k^2 + \alpha^2 k^4},
  \label{bog}
  \end{equation}
  where $|\psi_0|^2=|\hat \psi_{{\bf k=0}}|^2$ is the condensate density (which is $\approx N$ for a strong condensate).
  The Bogoliubov dispersion relation arises as a solvability condition for the equations resulting from the linearization of the GPE \eqref{GPE} with respect to a weak wave  with wavenumber $\bf k$.
  This dispersion relation is clearly observed in the  STFT in Fig.~\ref{fig:eos}(c), including the nonlinear frequency gap. Weaker branches on the bottom of Fig.~\ref{fig:eos}(c) appear because $\psi_{\bf k} $ is a  linear combination of the normal-mode amplitudes of the Bogoliubov modes ${\bf k} $ and $-{\bf k} $.)
  
  Like in the CB regime, the Fjørtoft dual cascade argument does not work in this regime. In the acoustic regime, there is no local inverse cascade, and most of $N$-flux is now toward the high $k$'s, whereas the low-$k$ range distribution is necessarily close to thermodynamic equilibrium (i.e., the role of the IR dissipation is negligible). The classical picture for such equilibrium is a condensate distribution 
  $N_0 \delta{(\k)}$ superimposed with thermal fluctuations in the form of the energy equipartition of the modes, the so-called Bogoliubov spectrum $ n_B(k) \propto T/k^2$ with constant $T$, which has the meaning of temperature. In an out-of-equilibrium three-wave regime (such as in the presence of a condensate), the flux $Q$ is proportional to the second power of $n_k$, and thus $T \sim |Q_0|^{1/2}$, which gives
  \begin{equation}
  n_B(k) \propto \frac {|Q_0|^{1/2}} {k^2}.
  \label{bogN}
  \end{equation}
  Note that the Bogoliubov $k$-scaling \eqref{bogN} coincides with the Rayleigh-Jeans energy equipartition scaling for the weakly interacting Bose gas governed by the four-wave mixing process.
  Finally, in a finite box, the condensate distribution takes the form of a "widened" $\delta$-function, i.e., a distribution sharply peaked around $k=0$ and rapidly decaying for $k\ne 0$.
  
  The presence of such a condensate corresponds to the short horizontal line at 
  $\omega \approx 1.34 \beta N$ in the STFT of Fig.~\ref{fig:st}(c), and enhanced in the inset of panel (e). Note that for pure uniform condensate, the STFT spectrum would be concentrated at a point  $(k,\omega) = (0, \beta N)$. In our case the condensate spectrum is spread in $k$, i.e. the condensate is non-uniform in physical space. Also, its frequency is shifted upward with respect to the uniform condensate case. Both of these features can be linked to the finite size of the computational box.
%
 One can take advantage of the fact that the wave and the condensate components in the large flux case are clearly separated on the STFT plot, and calculate the spectra of the waves and the condensate separately. To do that we band-pass filter the frequency spectrum in a narrow range  $1.30 \, \beta N <\omega < 1.38\,\beta N$ and call it condensate, while the rest of the spectrum we call waves. Then, performing the inverse Fourier transform in $\omega$ we get the $k$-spectra shown in Fig.~\ref{fig:st}(d), showing a thermodynamic Bogoliubov spectrum of waves over an extended $k$-range and a very peaked low-$k $ condensate's spectrum decaying approximately as $k^{-7}$.

Let us clarify the role of vortices in our system for different forcing intensities. Vortices are associated with incompressible kinetic energy, which arises from the Helmholtz decomposition of the regularized velocity field into solenoidal and potential parts.
 Fig.~\ref{fig:eos}(a) displays the energy components for various values of $|Q_0|$: incompressible and compressible kinetic energies, internal energy, and quantum energy.
 For low $|Q_0|$, quantum energy is dominant and internal energy is minimal, consistent with weak nonlinearity.
 Compressible and incompressible energies are equal, and each is half the quantum energy.
 The presence of incompressible kinetic energy indicates vortices, but these are "ghost" vortices without dynamical significance—they appear even in purely linear fields \cite{berry2000phase}. 
Interestingly, increasing $|Q_0|$ reduces incompressible kinetic energy to about 6–8\% in the strongest forcing cases, reflecting a sparse distribution of small vortex loops (with radii of the order of the healing length).
 This state contrasts with strong superfluid hydrodynamic turbulence states consisting of polarized tangles of fully formed quantized vortex lines of all sizes \cite{Polanco_VortexClusteringPolarisation_2021,Bradley_Fischer_QT_2025}. 
 In our system, strong vortex lines shrink and annihilate due to effective friction from the intense acoustic component. 
 Persistent vortex tangles would require sound damping, i.e., an effective \text{``cooling"} of the system (recall that the acoustic phonons correspond to a normal fluid component appearing at finite temperature). 
 Thus, the strong turbulence state reported here is dominated by interacting acoustic waves rather than dynamically significant hydrodynamic vortices.
A similar annihilation of vortices and transition to an acoustic regime was previously observed in 2D GPE simulations \cite{nazarenko2006wave}.
  
We will now  discuss the out-of-equilibrium equation of state  \cite{dogra2023universal,zhu2024turbulence}, which relates the global amplitude of the waveaction spectrum to the input flux, in our case $|Q_0|$. In practice, we fit our data with $n_k=A k^{-x}$ in the different observed ranges. The exponents $x$ were found to be very close to those discussed earlier, and the amplitude $A$ is strongly dependent on $Q_0$. Readers can find the fitting method and fitted exponents $x$ in  
Appendix \ref{append:c}.
Figure \ref{fig:eos}(b) shows the plot of the spectrum's amplitude $A$ vs. the particle flux in the inertial range $|Q_0|$ for our runs covering the whole range of particle fluxes, from low to high. 
For low-flux runs, we observe $A\sim |Q_0|^{1/3}$ in agreement with the weak four-wave WT. For intermediate fluxes, the spectrum at the high-$k$ part remains weak and still follows $A\sim |Q_0|^{1/3}$. The low-$k$ part, which exhibits the CB exponent, displays a dependence on the flux that could be fitted by $A\sim |Q_0|^{2/3}$. In other words, we obtain $y=2/3$ in the expression for the CB spectrum \eqref{eq:CB}. Such an exponent evokes some type of hydrodynamic Kolmogorov turbulence, but we lack a theoretical explanation and scaling range is too narrow for considering this exponent's value reliable.
Finally, for high-flux runs, we only measure $A$ for the large-$k$ thermodynamic part because the low-$k$ condensate spectrum range is too short for a reliable fitting. Here we see $A=T\sim |Q_0|^{1/2}$, which fulfills our prediction for the temperature scaling of the three-wave process. 

\Sec{Conclusions}
In this Letter, we studied the stationary inverse cascade states in BEC turbulence for a large range of forcing strengths (or equivalently particle fluxes). We showed that the predictions of the WWTT theory are very robust; namely, the KZ spectrum scaling is observed for fluxes spanning over more than four orders of magnitude.
At stronger fluxes, the waveaction spectrum and the STFT spectrum indicate the presence of the critical balance state in which the nonlinear frequency broadening $\delta \omega (k)$ is close to the linear wave frequency $\omega_k$ in a range of wavenumbers $k$. In the simulations with the strongest flux, we observe states close to low-temperature thermodynamic equilibria, with a prominent condensate component sharply peaked near $k=0$ and over-condensate fluctuations with spectra close to the thermal Bogoliubov spectrum $n_k \sim 1/k^2$. Our complete set of simulations has allowed us to determine a rich equation of state for the 3D inverse cascade BEC turbulence.
Experimentally, such a steady-state inverse cascade is yet to be implemented for 3D BEC turbulence.
As with a similar setup for the direct cascade \cite{navon2019synthetic}, the steady state will require simultaneous presence of forcing and dissipation.
However, now the dissipation has to withdraw low-momentum particles, and the forcing at high $k$ would have to provide new particles, which is not achievable by shaking the trap (a forcing used in \cite{navon2019synthetic}).
This could be done using a particle recirculation protocol 
devised in a recent experiment \cite{karailievObservationInverseTurbulentWave2024}: it provides effective dissipation at low-$k$ modes through particle removal, while simultaneously acting as a forcing term for particles at small scales by reinjecting them at the forcing scale $k_{\rm f}$.  Counting the number of removed and reinjected particles provides a natural measurement of the particle flux $Q$.
Clearly, experimental dissipation could not follow the exact shape used in our simulations, but it is likely that the details of the dissipation procedure are not crucial. The situation is similar with the synthetic dissipation implemented at high $k$ \cite{navon2019synthetic}, which was different from what is normally considered in the numerics of the direct cascade, but the results confirmed the theory in the inertial (non-dissipative) range of scales. This is consistent with the principle of turbulence universality, which states that 
the details of the forcing and dissipation are  unimportant in the inertial range --- it is only the spectral flux $Q$ that matters. Thus, we expect that the essential properties of the steady state will also be observed experimentally in the forced-dissipated inverse cascade settings.

\begin{acknowledgments}
This work was funded by the Simons Foundation Collaboration grant Wave Turbulence (Award ID 651471). 
This work was provided with computing HPC and storage resources of IDRIS and CINES under the allocations
A0152A14637 and A0172A14637 made by GENCI on the
Jean Zay (SKL partition) and Adastra (GENOA partition) supercomputers
This work was granted access to
the OPAL infrastructure from Université Côte d’Azur, supported
by the French government, through the UCAJEDI Investments in the Future project managed by the National Research
Agency (ANR) under Reference No. ANR-15-IDEX-01. 
\end{acknowledgments}

\section*{Data Availability Statement}
The source data used to generate all the figures in this study are openly available in the Figshare repository at the following URL: 10.6084/m9.figshare.31230085. The raw numerical data supporting the findings of this study are available from the corresponding author upon reasonable request, as the total data size exceeds the capacity for standard public archiving.

\bibliography{WWTT}
\appendix
  \renewcommand\thefigure{A\arabic{figure}}
  \renewcommand\thetable{A\arabic{table}}
  \setcounter{figure}{0}
  \setcounter{table}{0}

  \setcounter{equation}{0}
  \renewcommand\theequation{A\arabic{equation}}

\section{Numerical parameters}  \label{append:a}
\begin{table}[h]
\begin{ruledtabular}
\resizebox{\columnwidth}{!}{%
\begin{tabular}{|c|c|c|c|c|c|c|c|}
 case  &$f_0$  & $D_{\rm L}$  &  $r$ & $k_{\rm L}$ & $D_{\rm R}$&  $s$  & $D_0$ \\
\hline
0  &$10^{-4}$  & 1  & 0.5 & 6 & $4.29\times 10^{-26}$ & 6 & $10^3$ \\
1  &$2\times 10^{-4}$  & 2  & 0.5 & 6 & $2.54\times 10^{-30}$ & 7 & $10^3$ \\
2  &$4\times 10^{-4}$  & 2  & 0.25 & 7 & $1.50\times 10^{-34}$ & 8 & $10^3$ \\
3  &$8\times 10^{-4}$  & 4  & 0.25 & 8 & $8.89\times 10^{-39}$ & 9 & $10^3$ \\
4  &$1.6\times 10^{-3}$   & 8  & 0.25 & 8 & $5.26\times 10^{-43}$ & 10 & $10^3$ \\
5  &$3.2\times 10^{-3}$   & 10  & 0.25 & 8 & $3.11\times 10^{-47}$ & 11 & $10^3$ \\
6  &$1.2\times 10^{-2}$   & 16  & 0.25 & 8 & $1.63\times 10^{-55}$ & 13 & $10^3$ \\
7  &$2.5\times 10^{-2}$   & 70  & 0.5 & 10 & $1.63\times 10^{-55}$ & 13 & $10^3$ \\
8  &$5\times 10^{-2}$  & 250  & 0.75 & 11 & $9.75\times 10^{-63}$ & 15 & $10^4$ \\
9  & 0.2  & 16  & 0.125 & 8 & $9.75\times 10^{-63}$ & 15 & $10^4$ \\
10  & 0.5 & 20  & 0.125 & 8 & $9.75\times 10^{-63}$ & 15 & $10^4$ \\
\end{tabular}
}
\end{ruledtabular}
\caption{\label{tab:param}%
Parameters for GPE simulations. 
}
\end{table}
All essential numerical parameters for simulation runs are given in
Table~\ref{tab:param}. To systematically study the transition from weak wave turbulence to strong turbulence, we fix the size of the computing domain in Fourier space with $k_{\rm max}=173$, and the forcing position with $k_{\rm f}=125$ for all the runs.
The hyperviscous prefactor $D_{\rm R}$ and exponent $s$ are chosen to absorb the particle flux beyond the forcing range, preventing high-$k$ pileup while avoiding dissipating the inertial range of the inverse cascade. Using hyperviscosity in this way is typical for turbulence simulations to address the finite maximum wavenumber in the simulation, a constraint not present in the original physical system. Similarly, $D_{\rm L}$, $r$ and $D_0$ are chosen to maximize the $k$-space inertial range by providing controlled dissipation at the largest scales.

\section{Flux spectra}\label{append:b}
\begin{figure}[h]
    \centering
    \includegraphics[width=0.98\linewidth]{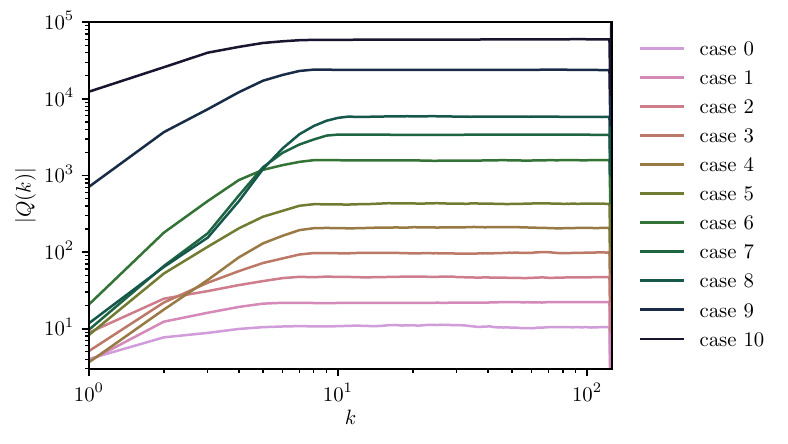}
    \caption{Particle flux spectra $|Q(k)|$. The characteristic flux $|Q_0|$ is determined from the quasi-constant plateaus observed for $k_\xi \le k \le k_{\rm{f}}$. The range $k > k_{\rm f}$ is omitted as $Q(k) \to 0$ due to particle conservation.
    }
    \label{fig:flux}
\end{figure}
The particle flux spectrum $Q(k)$ is computed as 
\begin{equation}
Q(k)=-2 \sum_{p=0}^k E_{\psi,\dot{\psi}}(p)\,,
\end{equation}
where $E_{f,g}(k)=\sum_{k-\Delta_k/2\le|\bm{k}|<k+\Delta_k/2} \hat{f}_{\bm{k}}\hat{g}_{\bm{k}}^*\,.$ and $\dot{\psi}=i\left[ \nabla^{2}  -|\psi|^{2} \right]\psi$ the Hamiltonian part on the RHS of the dimensionless GPE.
Figure~\ref{fig:flux} shows the flux spectra $|Q(k)|$ for all runs for $k\le k_{\rm f}$. 
The characteristic steady-state flux magnitude $|Q_0|$ is extracted from the well-defined plateaus in the range $k_\xi \le k \le k_{\rm f}$ and serves as the primary control parameter for the equation of state discussed in the main text.

\section{Method to generate the EoS figure}\label{append:c}
\begin{figure}[t]
\centering
\begin{minipage}[b]{0.48\textwidth}
    \centering
    \includegraphics[width=\linewidth]{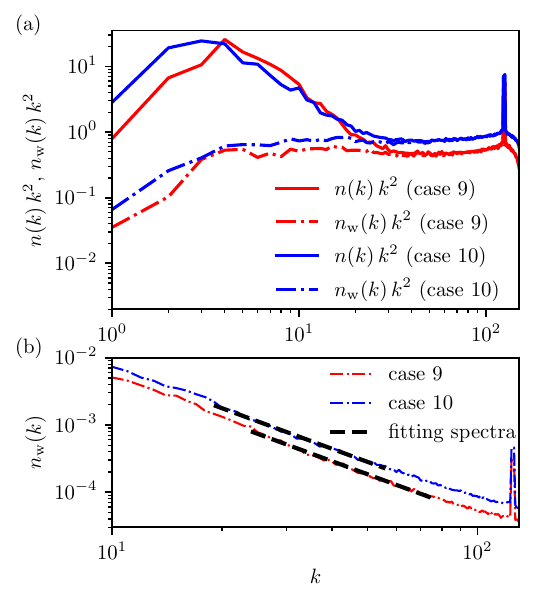}
    \vspace{-3mm}
    \caption{Power-law fits of the thermal components $n_{\rm w}(k)$ for high-flux cases. (a) Spectra compensated by the Bogoliubov scaling $n_{\text{B}} \sim k^{-2}$ to highlight the inertial range. (b) Power-law fits $Ak^{-x}$ superposed on the original spectra.}
    \label{fig:nkwfit}
\end{minipage}
\hfill
\begin{minipage}[b]{0.48\textwidth}
    \centering
    \includegraphics[width=\linewidth]{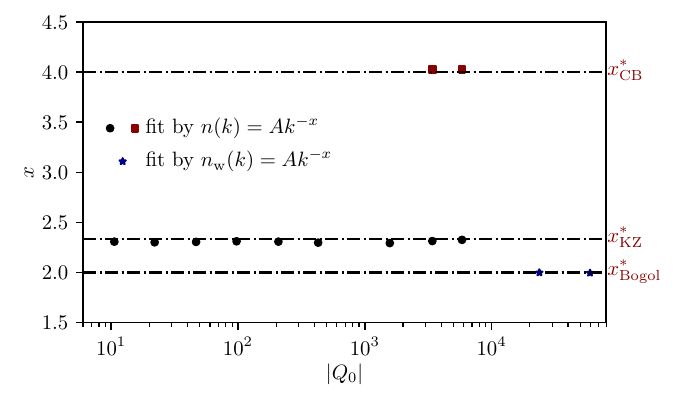}
    \caption{
    Fitted exponents $x$ corresponding to the equation of state in Fig.~\ref{fig:eos}(b). Theoretical predictions are shown for KZ ($x^*_{\rm KZ}=7/3$), Critical Balance ($x^*_{\rm CB}=4$), and Bogoliubov ($x^*_{\rm Bogol}=2$) regimes.
    }
    \label{fig:x}
    \vspace{-3mm}
\end{minipage}
\end{figure}
The out-of-equilibrium EoS is constructed by fitting numerical spectra to the power-law form $n = Ak^{-x}$. For high-flux cases 9 and 10, we fit the thermal wave component $n_{\rm w}(k)$, whereas for cases 0--8, the total spectrum $n(k)$ is used. This procedure is illustrated in Fig.~\ref{fig:nkwfit} for cases 9 and 10, where compensated spectra highlight quasi-constant plateaus in the $n_{\rm w}(k)k^2$ scaling. We extract the amplitude $A$ and exponent $x$ via linear regression of the logarithmic values, $\ln n = \ln A - x \ln k$, as shown in Fig.~\ref{fig:nkwfit}(b). This methodology is applied consistently to the weak turbulence ($k > 1/\xi$) and critical balance ($k < 1/\xi$) regimes. The resulting exponents $x$ are summarized in Fig.~\ref{fig:x} and correspond to the EoS presented in Fig.~\ref{fig:eos}(b).


\end{document}